\documentclass[lettersize,journal]{IEEEtran}
\usepackage{amsmath,amsfonts}
\usepackage{algorithmic}
\usepackage{algorithm}
\usepackage{array}
\usepackage[caption=false,font=normalsize,labelfont=sf,textfont=sf]{subfig}
\usepackage{textcomp}
\usepackage{stfloats}
\usepackage{url}
\usepackage{verbatim}
\usepackage{graphicx}
\usepackage{booktabs} 
\usepackage{cite}
\usepackage{ccicons}
\begin{document}

\title{Untangling Rhetoric, Pathos, and Aesthetics in Data Visualization}

\author{Verena Prantl, Torsten Möller,~\IEEEmembership{Senior Member,~IEEE,} and Laura Koesten}



\maketitle

\begin{abstract}
In contemporary discourse in data communication, logos (reason) and, more recently, ethos (credibility) have been discussed extensively. While the concept of pathos (emotional appeal) has enjoyed greater interest in the visualization community over the past few years, its connection to similar relevant concepts like rhetoric and aesthetics remains unexplored. In this paper, we provide working definitions of these terms, contextualizing them in data visualization, and explore their overlaps and differences in light of their historical development. Offering a historical perspective provides a more holistic understanding of how these approaches in science and philosophy have evolved over time, contributing to a better comprehension of their integration into the design process. Drawing from Campbell's seven circumstances, we illustrate how pathos is used as a rhetorical strategy in data visualizations today, at times inadvertently.

\end{abstract}

\begin{IEEEkeywords}
Human-centered computing, Visualization, Emotion/Affective computing, Visual rhetoric,
\end{IEEEkeywords}

\section{Motivation}
\label{section:1} 
In today’s data-driven world, the impact of data visualization goes far beyond presenting facts—it shapes narratives, influences decisions, and evokes emotions~\cite{park2022impact,lan2023affective}. While emotional appeal (pathos) is often underexplored in data communication, recent studies have begun to investigate its role, but its influence on the effectiveness of data visualization remains underestimated. Although the impact of aesthetics on visual displays has been researched, the findings have been somewhat inconclusive, demonstrating the need for further research~\cite{cawthon2007,thielsch2019experimental,lan2023affective}. However, research regarding emotions shows they are essential for connecting, understanding, and engaging with the world\cite{frijda1986emotions, lerner2015emotion}. In data visualization, where complex information is communicated visually, emotions are not to be neglected. They are integral to the sense-making process, potentially shaping how we interpret and engage with the data presented~\cite{dwyer2021struggling, Kennedy2018}. We believe that there is a need not just to acknowledge this aspect but to proactively address it. By doing so, we can improve how data is communicated and how people understand and respond to it~\cite{Heath2022book}. 
Researchers increasingly recognize the impact of design considerations beyond perception, emphasizing the importance of emotional goals in visualization. This shift reflects an understanding that visualization design and interpretation are influenced by more than what appears on the screen~\cite{akbaba2024entanglements}.
Recent studies, such as those by Akbaba et al.~\cite{akbaba2024entanglements}, Lee-Robbins \& Adar~\cite{lee-robbins_affective_2022}, Lan et al.~\cite{lan2023affective} or Matson et al.~\cite{Matson2018AffectiveDV}, reflect the growing interest in emotions within scientific discourse. However, a broader rhetorical and aesthetic framework for data visualization is still needed.
Our exploration of existing scholarly works regarding these key topics revealed several gaps that guided the direction of our research. First, there is limited research on emotional aspects in data visualization, specifically considering the broader rhetorical context.  
Secondly, although the historical evolution of rhetoric and aesthetics in data visualization has been explored~\cite{Kostelnick2016}, there is limited awareness within the scientific discourse on how the meanings of these concepts and their public acceptance have changed over time. We address this by providing a historical perspective on the evolution of rhetoric, pathos, and aesthetics to enhance understanding and inform current and future data visualization practices.
Furthermore, we offer an integration of interdisciplinary perspectives, as many studies tend to remain within their individual disciplines rather than adopting a holistic approach.

We draw on Kostelnick’s analysis~\cite{Kostelnick2016} of emotional appeals in data visualization through the lens of Campbell’s \textit{Philosophy of Rhetoric}~\cite{Campbell2008}, as well as Campbell \& Offenhuber’s~\cite{Campbell2019} subsequent exploration of this connection. We extend this research by offering a broad corpus of theoretical approaches connected to visual examples (in Section~\ref{section:circumstances}) that demonstrate how rhetorical strategies are embedded in data visualization design. This theoretical foundation serves as a starting point for our argument, urging a more conscious awareness of how emotional persuasion operates in data design. Since emotional persuasion in data visualizations may be used unintentionally by designers, it can lead to unintended consequences such as deception or a loss of credibility and effectiveness~\cite{cairo2019charts}.

This work appeals to the visualization community to engage with alternative theoretical foundations, incorporating etymological (examining the origins and evolution of concepts and terminology) and epistemological (exploring how knowledge is formed and validated) perspectives. By combining these approaches, we aim to untangle the complex influences of historical, cultural, and disciplinary factors as they evolve and offer a critical rethinking of established beliefs and concepts. It challenges existing assumptions and re-evaluates ideas concerning emotional appeals and aesthetics in the context of data visualization. The central question driving this research is: How can data visualizations be contextualized within the historical evolution of rhetoric, pathos, and aesthetics, and in what ways can this inform design decisions for data visualizations?
The contributions of this paper include working definitions for rhetoric, pathos, and aesthetics in the context of data visualization. By applying an etymological and epistemological perspective, we trace the evolution of these concepts and highlight their relevance to current practices. This paper also addresses a research gap by summarizing various scientific viewpoints on affective elements in data visualization. Drawing on various academic fields, including computer science, social sciences, cognitive science, psychology, philosophy, economics, and humanities, we approach the idea of rhetoric, pathos, and aesthetics in data visualization based on an extensive corpus of diverse scientific literature synthesized in this work. This establishes a foundation for understanding the intersection of visual rhetoric, emotional appeal, and aesthetics, enabling designers to integrate these considerations more consciously into their work. Moreover, it encourages designers to reflect on whether they already incorporate some of these elements subconsciously in their visualizations.

We argue that the role of pathos within rhetoric and aesthetics in data visualization has not yet been clearly articulated. To broaden the current discussion, we explore these terms in greater detail in Section~\ref{section:def}. We discuss the rhetorical triangle (logos, ethos, and pathos) in the Section~\ref{section:3modes}, and provide working definitions for these concepts within the context of data visualization in Section~\ref{sec:workingdef}. Furthermore, we focus on the historical embedding of rhetoric, pathos, and aesthetics, which we examine in Section~\ref{section:history}, and we outline the development of frameworks for visual rhetoric in Section~\ref{section:framework}. 
In Section~\ref{section:conncetion}, we examine how rhetoric, pathos, and aesthetics connect to each other as well as within a visualization context. We also discuss approaches that challenge the presumed neutrality of data, highlighting the pros and cons of emotion in visualization in Section~\ref{section:neutrality}. Furthermore, in Section~\ref{section:circumstances}, we adapt Campbell's framework from 1776 as first proposed by Kostelnick in 2016~\cite{Kostelnick2016} to the context of data visualization, using it to demonstrate how pathos techniques are actively employed in contemporary visualizations, supported by six examples. Finally, we offer design suggestions based on the literature discussed in the paper in Section~\ref{section:design_suggestions} and discuss our findings, along with their limitations and conclusions, in Sections~\ref{section:limitations} and~\ref{section:conclusion}.


\section{Related Work}
Aristotle defined rhetoric as the art of persuasion, forming the foundation for effective communication~\cite{crampton2001, Aristoteles, Campbell2019}. This concept applies not only to spoken or written language but also to visual channels, such as persuasive cartography. Persuasive maps convey geographic information while influencing opinions or behavior by selectively emphasizing certain details or omitting others~\cite{muehlenhaus2013design, muehlenhaus2012beyond}. Although often seen as manipulations, recent work in this area argues that maps should not only be judged by scientific standards but by how effectively they communicate the intended message. Maps are inherently rhetorical tools; while some prioritize accuracy and clarity, others are designed to persuade by shaping interpretation through specific design choices~\cite{muehlenhaus2013design, muehlenhaus2012beyond}. 
Effective persuasion, however, requires that the audience can comprehend and engage with the underlying concepts or patterns~\cite{wiggins2005understanding}. This consideration leads to the field of learning theory, which offers insights into how individuals process and understand information. Bloom’s taxonomy, introduced in 1956, provides a foundational framework for understanding learning processes. A more practical interpretation of this framework is offered by Wiggins \& McTighe,  outlining six facets of understanding: \textit{Explain, interpret, apply, have perspective, empathize, and have self-knowledge}. It is particularly interesting that according to these facets, empathy provides a basis for understanding~\cite{wiggins2005understanding}. It can be argued that explaining and interpreting align with logos as they involve logical analysis and reasoning. Ethos, however, relates to self-awareness and adopting a particular perspective due to ethical and moral dimensions in learning. Pathos should make the speaker's ethos credible while also emotionally connecting the audience to the logos of the speech~\cite{Herding2004, Busch2007, Zumbusch2010}.

Prior work in the area of visualization also addresses affective dimensions and visual rhetoric~\cite{lee-robbins_affective_2022, Kostelnick2008, Kostelnick2016, Campbell2019, Kennedy2018, lan2023affective, DIgnazioCatherine2020Df}. Lan et al., for example, systematically analyze projects and literature on affective visualizations and collect arguments around affective design~\cite{lan2023affective}. A study by Lee-Robbins \& Adar underscores the value of considering affective learning objectives in communicative visualizations. It highlights the importance of incorporating emotions to influence behavior through visualizations, sparking discussions on emotional impact in modern visual communication~\cite{lee-robbins_affective_2022}. Lee-Robbins \& Adar's taxonomy provides a solid foundation. Still, we argue for a more detailed exploration of rhetorical strategies in visualizations, including working definitions within the context of data visualization. 
A deeper understanding of these strategies, as we aim for in this paper, allows data visualization designers to tailor their visualizations to different audiences and contexts. 

The Tactical Technology Collective takes another interesting, more practical approach to these concepts. They explore how activists incorporate visual elements into their campaigns, with over 60 global examples demonstrating how visual information campaigns capture attention, convey narratives, and guide audiences through data-driven storytelling~\cite{visualpersuasion}.

Related work in visualization research has begun to explore frameworks that account for context, subjectivity, and the role of human perspectives. For example, Akbaba et al. argue for integrating feminist epistemological theories into visualization design. These theories highlight how knowledge creation is shaped by historical and contextual factors, suggesting that both data and visualizations, along with their designers and audiences, are inherently interconnected~\cite{akbaba2024entanglements}. Similarly, researchers in the area of ``Data Feminism,''~\cite{DIgnazioCatherine2020Df} argue that knowledge can emerge in various forms and that emotions can help better understand and remember data. Data Feminism recognizes visualizations as a form of rhetoric, addressing framing effects and critically examining the perceived neutrality of data~\cite{DIgnazioCatherine2020Df}. 
Our approach aligns with this shift, expanding the lens to include rhetorical, aesthetic, and emotive dimensions, focusing on how elements like rhetoric and pathos can enrich the interpretive experience. This paper uses a different theoretical framework grounded in Aristotle's rhetorical principles. We analyze the persuasive power of data visualizations, drawing on interdisciplinary research and highlighting their historical context. This approach shifts the analysis from a technical focus to a more theoretically grounded exploration of how data visualization may serve as a tool for persuasion within a rhetorical context.


\section{Methodology}

Navigating the complexities of literature syntheses that weave together a wide range of perspectives from different scientific fields, like social sciences, economics, humanities, and computer science, poses a methodological challenge. Our primary goal was to explore a broad spectrum of research in different disciplines, with a particular interest in incorporating heterogeneous sources~\cite{GREENHALGH2005417}. To address this, we identified the multidisciplinary meta-narrative review standard practices presented by Greenhalgh et al.~\cite{GREENHALGH2005417} as a fitting methodological framework. 

\subsection{Framework for Literature Review}

We used the framework proposed by Greenhalgh et al.~\cite{GREENHALGH2005417} to navigate and assess relevant literature across diverse scientific domains, even without conducting a complete systematic review. Their meta-narrative review process comprises six key phases: \textit{Planning, searching, mapping, appraisal, synthesis,} and \textit{recommendations}. Following this framework, we discussed the research question with a multidisciplinary team and used an open-ended format to explore the historical evolution of rhetorical elements in data visualizations. The research question was: How can data visualizations be contextualized within the historical evolution of rhetoric, pathos, and aesthetics, and in what ways can this inform design decisions for data visualizations? 
This approach allows for a broad coverage of peer-reviewed articles, encompassing contemporary and historical perspectives on rhetorical strategies with a thematic focus on visual data representation. Throughout the search phase, we employed diverse queries, as detailed in the following subsection, to explore a range of literature across different scientific domains. As a team, we assessed the significance of identified studies and papers, aiming for comprehensive insights into their contributions to our research.

\subsection{Search Strategy}
\label{section:search} 
The following databases were used to identify relevant literature: \textit{IEEE Xplore, Google Scholar, ScienceDirect, ACM digital library}, and \textit{Scopus}.

Different keywords and search phrases were used to ensure that the results were not limited to one scientific discipline. To obtain relevant results from the social sciences and humanities, search queries with the terms ``rhetoric'', ``pathos'', and ``aesthetics'' (``pathos in data visualizations,'' ``data + pathos,'' ``data + emotion,'' ``visual rhetoric'', ``aesthetics in data visualization,'' ``aesthetics + rhetoric'') were used. In the area of data visualization, on the other hand, the keyword ``pathos'' was nearly without any useful results, which is why more umbrella variants and synonym lists were chosen as queries (``empathetic emotions in data vis,'' ``data-driven storytelling,'' ``emotional data visualization,'' ``rhetoric + emotion,'' ``aesthetics + emotion,'' ``rhetoric + visualization'').

We included peer-reviewed scientific articles connected to the context of data visualizations from the past 20 years. Incorporating older literature was essential to provide a historical perspective on the topic and recognizing the enduring significance of foundational thinkers such as George Campbell (1719–1796) and Aristotle (384-322 BC) in the realm of humanities. By doing so, we aimed to establish a theoretical foundation for our research, drawing connections between contemporary scientific articles and the enduring insights of these historical figures, thereby enriching the depth and breadth of our analysis.
We searched the specified search queries in the title, abstract, or keywords of peer-reviewed, full-text publications. This helped to map the diversity of perspectives and approaches of the literature, as suggested by Greenhalgh et al.~\cite{GREENHALGH2005417}. By citation chaining, we identified relevant literature and related resources for the multiple scientific domains synthesized in this paper. Also, we allowed other relevant search terms to be identified (``emotional framing data visualization,'' ``emotional design,'' and ``communicating numbers''). Furthermore, we searched for additional peer-reviewed articles referencing the papers obtained in the preceding stages. Any pertinent papers found were included in our compilation, forming the basis of this work.

\begin{figure}[t]
	\centering
	\setlength{\fboxsep}{1pt}
	\setlength{\fboxrule}{1pt}
	
\includegraphics[width= 260pt]{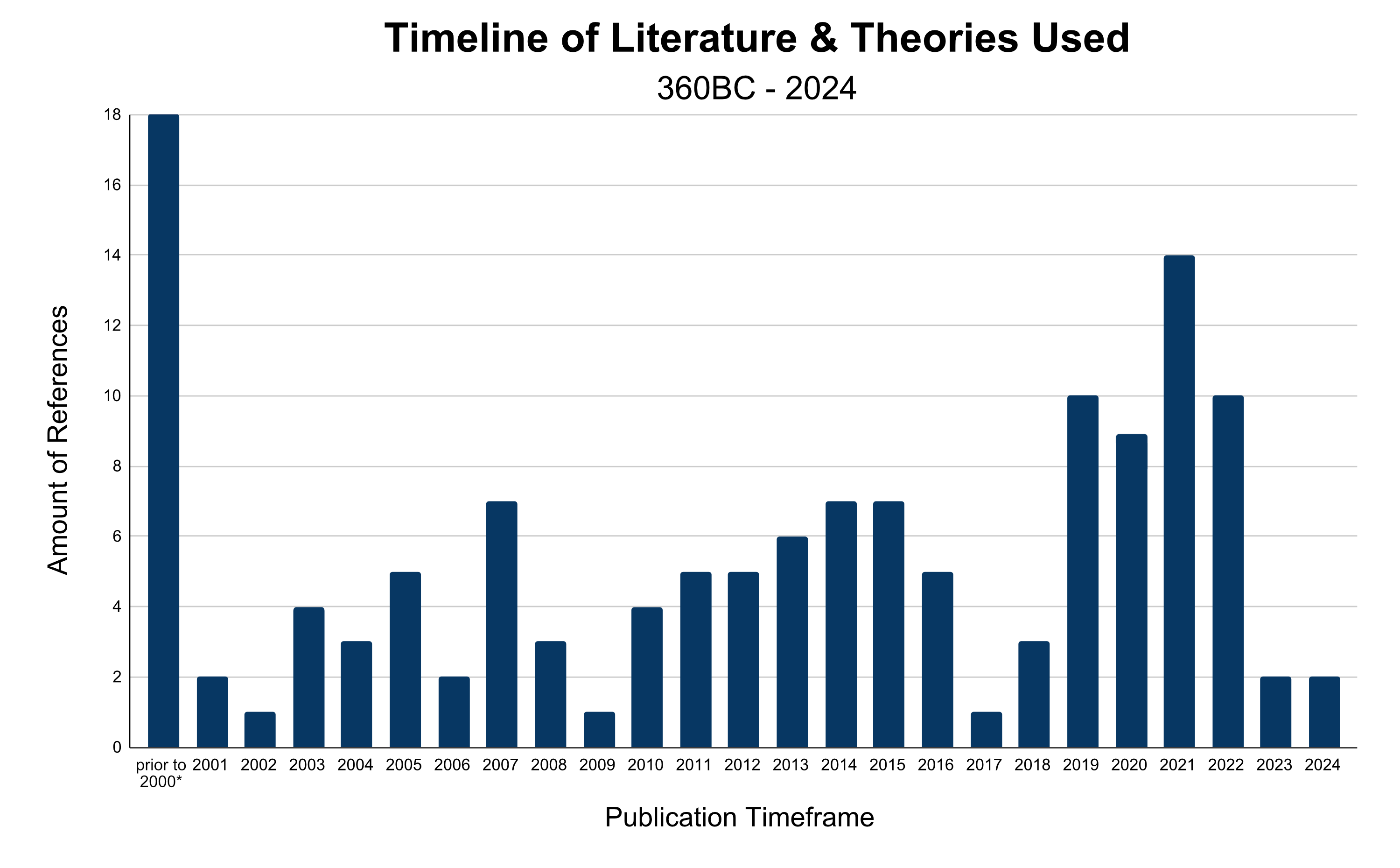}
	\caption{This chart visualizes the progression of cited works in the paper, providing an overview of the historical context and distribution of references to clarify the paper's temporal citation landscape. Visualization examples were excluded, as this chart focuses on the theories and literature used for argumentation and analysis. *References prior to 2000 are widely spread, the exact years being: 360 BC, 1776, 1777, 1937, 1948, 1972, 1976, 1985, 1986, 1987, 1988, 1990, 1991, 1997, 1998, 1999 (3x).}
	
	\label{fig:refernces}
\end{figure}

\subsection{Screening Process}
Adopting an exploratory approach, our methodology offers flexibility in navigating the complexities of a broad research domain. This non-systematic literature review is informed by iterative discussions to assess the eligibility of cited papers in the context of our research aims. Including diverse methodologies, we focused on work aligning with our central research theme — rhetoric, pathos, and aesthetics in contemporary and historical data visualizations.
Our choice of this exploratory approach was driven by the need for flexibility and adaptability in navigating the complexities of our research domain, which encompasses a wide spectrum of disciplines and diverse methodologies.
Papers focusing on trust, casual visualization, design principles in general, guidelines, and current literature about rhetoric without a focus on pathos or visualizations were excluded. These exclusions were made to maintain the focus and relevance of our paper. By focusing solely on rhetoric, pathos, and aesthetics in data visualizations, we aimed for a nuanced examination of the relationship between rhetorical elements and data visualization, thereby enhancing the quality and coherence of our research.

The chronological timeline presented in Figure~\ref{fig:refernces} serves as a navigational aid for readers, offering a visual depiction of the literature and theories used throughout this paper that show the historical evolution of rhetorical, aesthetic, and affective dynamics in data visualizations. By highlighting patterns and shifts in the distribution of cited literature and theories over time, Figure~\ref{fig:refernces} supports both an etymological approach---examining the origins and evolution of key concepts and terminology---and an epistemological perspective, which explores how knowledge in the field has been formed and validated across different periods.

\section{Rhetoric, Pathos, and Aesthetics in Data Visualization}
\label{section:def} 
As a first step, our objective is to explore different definitions of rhetoric, pathos, and aesthetics, connecting them to data visualization and then creating working definitions (in Section~\ref{sec:workingdef}) that will guide the rest of the paper and serve as a foundation for the discussions and analyses that follow.

\textbf{Rhetoric} describes the art of persuading and influencing others by using language~\cite{Aristoteles, KinrossRobin1985TRoN}. It is one of the seven liberal arts and emerged in the early fifth century BC. It originally focused on persuasive public speaking, but it has expanded to include written communication~\cite{LanhamRichardA1991Ahor}. Throughout its history, it has oscillated between specific technique-oriented concerns and broader ethical considerations, leading to varied definitions. Aristotle, for example, defined three kinds of rhetoric because this is ``how many different kinds of audience there are for speeches,''~\cite{Aristoteles} namely: deliberative (to recommend or warn against something, often used in political contexts), judicial (to accuse or defend somebody, often used in legal contexts), and epideictic (to praise or blame someone, often used in ceremonial contexts)~\cite{Aristoteles}. Aristotle is considered the founder of the rhetorical triangle (further explained in Section~\ref{section:3modes}), with logos (logic), ethos (ethical concerns), and pathos (emotional appeal) as cornerstones~\cite{CorbettEdwardP.J1999Crft}. Cicero, however, defines three different functions for an orator, and thus for rhetoric: To teach, to please, and to move~\cite{LanhamRichardA1991Ahor}. Universal rhetoric, which explores the power of language across all discourse, establishes rhetorical studies as a key intellectual discipline~\cite{JasinskiJamesL2001SoR:}. Campbell \& Offenhuber define rhetoric in an Aristotelian sense as the study of persuasion~\cite{Campbell2019}. Although they acknowledge the controversy surrounding the discourse about persuasion in data visualization, they assert that the presentation and context of visualizations often communicate as much as the data themselves~\cite{Campbell2019}. This shows that defining rhetoric's scope has proven challenging, exemplified by the overlapping domains of persuasion, such as philosophy, leading to enduring debates~\cite{LanhamRichardA1991Ahor}. Nonetheless, rhetoric is a way of conveying ideas, emotions, and intentions. It is the skill of adapting language to its purpose, making rhetoric a versatile concept~\cite{weinsheimer}. Rhetoric can extend beyond language and be applied to different media. Today, there are emerging frameworks for visual rhetoric (see Section~\ref{section:framework}).

\textbf{Pathos} is one part of rhetoric and encompasses techniques for emotional appeal, referring to both the speaker's and the audience's emotions~\cite{CorbettEdwardP.J1999Crft, Aristoteles}. Aristotle understood emotional appeals (pathos) as being based on ``pain and pleasure'' and categorized emotions into opposites, such as anger versus calmness, friendship versus hostility, and fear versus confidence~\cite{KostelnickCharles2019Hvd:, Aristoteles}. He described these emotions as ``those affections, accompanied by either pain or pleasure, that cause people to undergo a change and modify their judgments''~\cite{Aristoteles} and linked them to tragic emotions and catharsis in the ancient theater~\cite{UedingGert1976EidR}. By evoking emotions alongside logical arguments and evidence, a speaker can effectively persuade an audience~\cite{KostelnickCharles2019Hvd:, Aristoteles}. Still, attempts to define pathos often lack clarity and rigor~\cite{rees_1972}. Terms such as empathetic emotions and emotional appeals can be used as an approximation to pathos and narrow down emotions to those that ``affect a person's soul.''~\cite{Gessmann2002, rees_1972} Because of its transforming meaning throughout history (further described in Section~\ref{section:history}), finding a unified description of pathos in relevant literature remains difficult~\cite{rees_1972, SulzerJohannGeorg1777ATds}. As Campbell \& Offenhuber note, emotional appeals in visualization are often associated with visual embellishments, which are frequently debated (see Sections~\ref{section:conncetion} and~\ref{section:neutrality})~\cite{Campbell2019}. Within the visualization literature, there are ongoing discussions about the designer's intention regarding these kinds of emotional appeals~\cite{lee-robbins_affective_2022} and how they could be used to get a targeted affective response~\cite{Matson2018AffectiveDV}.

\textbf{Aesthetics} is considered a branch of philosophy that deals with the nature of beauty, art, and taste. The term originates from the Greek word \textit{aisthesis}, which translates to ``sense perception'' or ``sensory cognition.''~\cite{CarrollNoel1999Poa:} Since the time of philosophers like G.W.F. Hegel, aesthetics has been explored as a theory of the beautiful, examining how art is perceived and valued. Hegel's contributions emphasized the importance of art in expressing ideas and reflecting societal values, which have shaped contemporary understandings of aesthetics~\cite{HegelGeorgWilhelmFriedrich1998A:lo}. In the $18^{th}$ century, the concept of aesthetics expanded to encompass themes such as the sublime, magnificent, elegant, graceful, as well as the ugly and grotesque, forming a comprehensive theory of aesthetic values, their experience, and evaluation~\cite{KutscheraFranzvon2019Ä}. In contrast to traditional beauty-centric views, empirical aesthetics focuses on intense emotions and interactions with objects. It includes research on beauty, aesthetic pleasure, preferences, art perception, evaluation, and creation~\cite{brielmann2018aesthetics}. Originating in the $19^{th}$ century with Gustav Fechner, empirical aesthetics has received interest from various fields, such as neuroscience, psychology, and the wider public~\cite{brielmann2018aesthetics}. This has led to defining aesthetics as the study ``of how and why sensory stimuli acquire hedonic value.''~\cite{skov2020farewell}
Aesthetic experiences are closely connected to the stimulation of the brain and the experience of pleasure. Additionally, evolutionary factors, such as symmetry and attractiveness, inform aesthetic preferences, serving as indicators of health and mate quality~\cite{brielmann2018aesthetics}. In data visualization, aesthetics refers to the design elements—such as color, layout, and typography. Aesthetics should balance beauty with function, improving both the viewer’s experience and understanding of the data.

\subsection{The Rhetorical Triangle: Logos, Ethos, and Pathos}
\label{section:3modes} 
As previously mentioned, the foundation of Aristotelian rhetoric is built on logos, ethos, and pathos as its cornerstones~\cite{Aristoteles}. These three modes of persuasion have different values within science but need to be balanced to achieve effective communication~\cite{Aristoteles, Campbell2008}. Therefore, each mode is introduced briefly and contextualized into the field of data visualization; however, pathos remains central to this work, as it has not been explored in visualization research to the same extent as logos and ethos.

\textbf{The concept of logos} can be referred to as clarity (e.g., \cite{Tufte1990, Tufte2006, Tufte2007}). Clarity is the principle stating that data visualizations should be optimized for perception and the task they support. It is mainly about transmitting fact and truth~\cite{brasseur2003visualizing}. Logos as a form of clarity is already highly valued in the visualization community ~\cite{Kostelnick2008, Tufte2006, Tufte1990, Tufte2007}. 

\textbf{The concept of ethos} is a compendium of aspects extending beyond classical ethics. Ethos in this context includes the credibility of the source, but also the adaptation of the visualization to the target group (social rhetoric), as well as interactive functionalities that allow viewers to change the chart and explore it on an individual level~\cite{Kostelnick2008, Correll2019}. Ethos has only recently been discussed as an issue not to be neglected in data visualization research~\cite{Correll2019}. Data or data visualizations are often considered apolitical and ethically neutral, as data itself are seemingly objective, which can be seen as dangerous~\cite{Correll2019, boehnert2016data} and which will further be discussed in Section~\ref{section:neutrality}. Correll~\cite{Correll2019} states that people creating data visualizations also have moral obligations to take the ethical dimensions of the particular visual representation of data into account, even if that means they are sometimes confronted with ethical dilemmas.

\textbf{The concept of pathos} is the appeal to an audience's emotions. The goal is not only to convey factual information but also to evoke a sense of empathy within the recipients, making them feel as though the expressed concern is their own~\cite{Herding2004, Busch2007, Zumbusch2010}. This theory is also supported by studies in neuroscience, showing that an empathetic response to someone else's experience, ``involve[s] the same brain tissue that's active when you yourself have that experience.''~\cite{Bloom2016} This aligns with Aristotle’s concept of catharsis, which he described as ``a moment of renewal.''~\cite{Aristoteles}

\subsection{Working Definitions of Rhetoric, Pathos, and Aesthetics in Data Visualization}
\label{sec:workingdef}

For the purposes of this paper, we propose the following working definitions of rhetoric, pathos, and aesthetics in the context of data visualization.

\textbf{Working definition of rhetoric:} Rhetoric is the overarching framework that guides the use of communication techniques to persuade or engage an audience. In data visualization, this means using visual elements to effectively communicate data by appealing to the audience’s logic (logos), credibility (ethos), and emotions (pathos). This approach integrates aesthetic and symbolic aspects to present information clearly while also engaging the viewer.

\textbf{Working definition of pathos:} Pathos is a key part of rhetoric, focusing specifically on emotional appeal. In this paper, the term pathos describes visual elements used to evoke emotional responses in an audience. This involves creating connections that resonate emotionally with the viewer, making data more relatable and understandable without resorting to manipulation.

\textbf{Working definition of aesthetics:} Aesthetics refers to the visual appeal and beauty of things. In data visualization, aesthetics entails visual design choices, such as color, symmetry, and composition, that can support rhetorical goals by making the information more engaging and compelling.

In the following section, we use an etymological (examining the origins and evolution of concepts and terminology) approach to explore the historical development of rhetoric in general, with a particular emphasis on the evolution of pathos, and discuss the emergence of visual rhetoric.

\section{Historical Context of Rhetoric \& Pathos}
\label{section:history} 
Rhetoric has undergone different connotations throughout its history. In ancient times, rhetoric was important because adult, free male citizens in Greece could participate in political and legal decisions but had to speak publicly to do so~\cite{raaflaub013citizenship}. Aristotle was among the first to develop a systematic representation of the art of oratory (i.e., formal speeches that strongly affect people's feelings). He defined it as the ability to consider what might be persuasive in any matter~\cite{raaflaub013citizenship, Aristoteles}.
Rhetoric was used throughout history as an effort to legitimize power in relation to monarchical orders~\cite{MeisterJanB.2021AumP}. Having said that, Bonsiepe's argument on rhetoric should be taken into account, as he stated that: ``Where there is force, rhetoric is not needed; in fact, rhetoric cannot exist under force. Persuasion relies on the ability to choose.''~\cite{Bonsiepe2008} 

During the Age of Enlightenment, rhetoric was criticized for diverting attention from rational thinking. This criticism grew stronger in the late $18^{th}$ century with the rise of the aesthetics of genius. Speeches were valued for their emotional authenticity rather than skillful techniques~\cite{KantImmanuel1948EzKd, ueding1997goethes}. Rhetoric faced moral scrutiny as a manipulative ``art of dissimulation,'' resulting in its decline as a teaching subject in the $19^{th}$ century. Even Goethe, who had received rhetorical training, disapproved of it~\cite{KantImmanuel1948EzKd, ueding1997goethes}.
Karl Marx then introduced a new meaning to rhetoric, especially in communism. A socialist rhetoric with special terminology evolved, which interpreted political conditions in the sense of Marxism and supported the argumentation of its representatives. However, this resulted in the civil population viewing rhetoric speeches negatively~\cite{PlewniaAlbrecht2011UF&A}.

Interestingly, the historical background of pathos is somewhat different. At the end of the $18^{th}$ century, Johann Georg Sulzer probably marked the peak of the pathos-theory within the fine arts when he described pathos as the nature of great souls in his book ``Allgemeine Theorie der schönen Künste'' (engl. ``A General Theory of the Beautiful Arts'') ~\cite{SulzerJohannGeorg1777ATds}. 
Only in the course of the $19^{th}$ century the importance of pathos started to decrease, and in the $20^{th}$ century, the term was increasingly associated with exaggeration, kitsch, and triviality~\cite{Zumbusch2010}. A similar trend can be noted in data visualization, where emotional appeal experienced a comparable rise and decline.

Kostelnick provides a historical perspective of using pathos in data visualizations~\cite{Kostelnick2016}. He calls the $19^{th}$ century the ``golden age of statistical graphics'', a term he borrowed from H. Gray Funkhouser~\cite{Funkhouser1937}. Victorian and romantic values dominated in this so-called ``golden age'', and these values are reflected in the design of contemporary data visualizations~\cite{Kostelnick2016}.
Around this time, in the second half of the $19^{th}$ century, new graphic forms emerged, including Minard's map (see Figure~\ref{fig:examples}, E), visualizing Napoleon's Russian campaign in 1878. Quantitative information was presented in new media, such as US statistical atlases and French national statistics albums, to name a few. This not only made data visualizations accessible to a wider public but also created emotions by using a combination of physical and psychological stimuli, including lavish color, pictorial elements, and innovative genres~\cite{Kostelnick2016}.

Similar to the historical development of pathos in the field of fine arts and rhetoric overall, the acceptance of appealing to emotions started to decrease in data visualizations in the $20^{th}$ century. Kostelnick sees an explanation of this avoidance of pathos in the rise of a new modernist minimalism---emotions had no place in serious data visualizations. However, in recent years, a more open attitude towards using pathos in culture, politics, and even science and data visualizations is notable~\cite{Kostelnick2016, Dachselt2003, Hayer2021}. 
Rhetoric, in general, experienced a renaissance during the $20^{th}$ century. A renewed interest in Europe and North America led to a ``return of rhetoric''. Nietzsche and Heidegger, in particular, have inspired other and following philosophers and theorists to return to the debate on rhetoric~\cite{AnschützHans-Peter2021NR, 2009Hkopperschmidt}. Contemporary rhetoric encompasses a variety of concepts, including the relatively new field of visual rhetoric, which extends traditional persuasive techniques to visual media. This differentiation is important because words and images are processed differently: words are read sequentially, while images can be understood sequentially and simultaneously. For instance, 3D objects can be projected onto flat surfaces in maps, and images can be layered with latitude lines. Additionally, visual sequences can be animated to show time progression in films and cartoons~\cite{gross2013science}. Today, many posters, ads, films, and TV commercials include both text and images~\cite{PeninBonsiepe2022VR1}. The next section discusses the concept of visual rhetoric and the emerging frameworks for rhetoric in visualization.

\subsection{Frameworks for Visual Rhetoric}
\label{section:framework} 
Visual rhetoric starts with a designer crafting visuals tailored to a specific audience and purpose and is subsequently interpreted by the audience within a particular context. However, this action is influenced by larger social factors. Visual language is shaped by communities that train their members in its conventions, which reflect cultural values, norms, and aesthetics~\cite{KostelnickCharles2004MIMA}. Icons, symbols, and colors can carry different meanings in various cultures, posing challenges and risks for creating information designs aimed at international or multicultural audiences~\cite{kostelnick2003shaping}.


In the first half of the $20^{th}$ century in the United States, rhetoric mainly focused on spoken communication, guided by classical rhetoric principles. The emphasis was on verbal aspects, with less attention to visuals. However, the second half of the $20^{th}$ century witnessed significant changes. The rise of television made visual persuasion more prominent. Political movements started using ``photo opportunities,'' and politicians became skilled in ``image management.''~\cite{kenney2003review} Philosopher Kenneth Burke's conceptualization of rhetoric broadened its scope to include nonverbal forms of communication, impacting various academic disciplines. Semiotics (the science of signs and symbols and their processes of development in culture and nature) often considers pictures as ``iconic'' or ``indexical'' signs. These terms suggest that pictures have a natural or direct link to what they represent. For example, an iconic sign, like a drawing of a tree, resembles a real tree, while an indexical sign, like smoke, points to a fire as its cause.

In contrast, a rhetorical perspective views symbols differently. It assumes that symbols in language and images are ``conventional,'' meaning they rely on social agreement. Rather than having a natural connection to what they represent, symbols are considered arbitrary---they depend on a shared cultural understanding rather than any inherent resemblance or causal link to their meaning.~\cite{kenney2003review, lizardo2016cultural}.

In the postmodern era, visual design is viewed as a language connecting designers and audiences, particularly in architecture. Tailoring a design to a community’s social and cultural context evokes emotion by resonating with the audience’s identity, making the visuals feel more personal~\cite{KostelnickCharles2019Hvd:}. Social factors are inherently persuasive, influencing how communities use visual language to achieve their objectives, as seen in $19^{th}$ century America with the emergence of data displays for presenting statistical information~\cite{KostelnickCharles2004MIMA}.
This kind of rhetoric, whether verbal or visual, showcases shared values and reinforces common beliefs. Pictures, in particular, have the power to make people feel like they are experiencing events firsthand~\cite{Kjeldsen2021}.

For instance, in 2021, Pflaeging examined the rhetorical impact of the fusion of layout, typography, and color in magazine design, emphasizing the crucial role of visual structure as it underscores and provides guidance, signaling what is significant on a magazine page and what is not~\cite{pflaeging}. This approach aligns with the work of Hullmann \& Diakopoulos~\cite{Hullman2011}, who identified the following classes of rhetoric for visualization, further illustrating how design elements can influence understanding and engagement:
\begin{itemize}
\item \textit{Information Access Rhetoric}, including omission to simplify complex ideas, using metonymy (a linguistic expression used figuratively, not literally) to represent parts of a whole for simplification, using averaging and aggregation.
\item \textit{Provenance Rhetoric}, including signaling transparency and trustworthiness by linking data sources, effectively representing uncertainty, and providing information about the visualization designer.
\item \textit{Mapping Rhetoric}, including obscuring through noise and sizing changes, employing visual metaphors and metonymy for implicit meanings, and utilizing contrast, classification, and redundancy methods to enhance communication in data visualizations.
\item \textit{Linguistic-based Rhetoric}, including typographic emphases like font bolding and irony via rhetorical questions, showing similarity through analogies and metaphoric statements and individualization methods such as framing from an individual perspective.
\item \textit{Procedural Rhetoric}, including rule-based representations and interactive functions for conveying meanings, anchoring techniques like default views, fixed comparisons, or spatial ordering, and interactive methods such as individual filtering options through search bars or menu options~\cite{Hullman2011}.
\end{itemize}

Another relatively new approach to visual rhetoric is speculative visualization. Speculative design seeks to uncover hidden social issues in daily life by engaging in diverse design practices like participatory workshops and conceptual proposals. Speculative visualization is a subset that uses visual methods to encourage reflection, bringing together various fields like art, aesthetics, scientific data analysis, and humanitarian perspectives~\cite{kim2010speculative}.

In the next section, we will discuss how rhetoric and pathos are connected to aesthetics in data visualization.

\section{Interconnection of Rhetoric, Pathos, and Aesthetics in Visualization Design}
\label{section:conncetion} 
Rhetoric and aesthetics are two discourses that use different principles and frameworks to approach and understand objects, concepts, processes, and more. Aesthetics, in principle, deals with beauty and pleasure as sensual experiences~\cite{goldman2005aesthetic}. 
Aristotle views emotions (pathos) as integral to ethical behavior, arguing that they contribute to a person's ability to act virtuously. He also emphasizes aesthetics in actions and character, connecting beauty and harmony with a well-lived ethical life. Additionally, ethos (credibility) intersects with emotions and ethics, as the trustworthiness of one’s character often depends on how emotions are managed and expressed. There is a blurred line between emotional influence and ethical judgment, emphasizing that cultivating virtue involves balancing rationality, emotions, and aesthetics~\cite{milliken2006aristotle}. 

Cicero defined the five canons of rhetoric as the essential skills needed for effective and persuasive communication: \textit{inventio (finding a basic argument), dispositio (structure of the argument), elocutio (choosing words and phrases), memoria (content memorization)} and \textit{actio (effective delivery like tone, voice and expression)}~\cite{Schirren+2008+620+630}. The connection between aesthetics and rhetoric mainly stems from  \textit{elocutio}, involving stylistic devices and linguistic composition. Rhetoric uses aesthetic principles as a practical tool, while aesthetics introduces creative freedom into the rhetorical practice~\cite{Rüsen+2005+41+58}. To put it differently, rhetoricians leverage aesthetic principles to enhance the persuasiveness of their arguments, assigning a purpose to aesthetics by incorporating those principles into their communication. Aesthetics governs the foundational principles of visual appeal and beauty, which are inherently integrated into the design process~\cite{KnapeJoachim2021PbDD, burdek2005design}. Ultimately, design serves as the practical application of these aesthetic principles, allowing for creating visually captivating objects, spaces, and means of communication. The term ``design'' originates from the Latin word ``designare,'' which initially means ``to designate'' and evolved to denote ``to draft'' during the $15^{th}$ century~\cite{KnapeJoachim2021PbDD, burdek2005design}.
Within design methodology, the primary focus lies on expressive elements during the drafting process, ultimately shaping objects and structures~\cite{KnapeJoachim2021PbDD, burdek2005design}. 
Tufte introduced the term ``chartjunk'' into this discourse to critique content-free embellishments in data visualizations~\cite{Cairo2013}. Tufte’s aversion to ``chartjunk'' originates from the suspicion that promoters of this type of decoration view facts and figures as boring, which then require embellishments to bring them to life. However, Tufte states that ``if the numbers are boring, then you've got the wrong numbers. [...] Who would trust a chart that looks like a videogame?''~\cite{Tufte1990}.

However, Tufte's concept of ``chartjunk'' encounters resistance in today's academic visualization community. This opposition stems from concerns about labeling essential icons as ``chartjunk,'' which is perceived as inaccurate and detrimental to scientific discourse.
In their ``Manifesto for Putting `Chartjunk' in the Trash 2021,'' Abkaba et al. explore the term's history. They argue that the term was originally meant to provoke and devalue design approaches in visualizations, ultimately promoting Tufte's minimalist approach as the ``way to truth''~\cite{Akbaba2021}. 
Hullman et al. also challenge the conventional idea of making visualizations efficient for quick understanding. Instead, they suggest that introducing visual difficulties can improve comprehension and recall~\cite{hullman2011benefitting}. 

This argument is underlined by Norman's ``Emotional Design'' concept~\cite{NormanDonaldA2003Tdoe}. The main rationale for his theory is simple: Beautiful things are also more functional~\cite{Cairo2013}.


However, debates continue regarding the appropriateness of aesthetics in data visualization, including what these concepts entail and the purposes they should serve. This is highlighted in a discussion from Stephen Few with Mike Danziger, where Few emphasizes the importance of creating engagement in visualizations. He defines engagement as immersing viewers in exploring and comprehending information without being distracted by software mechanics. Few emphasizes that the primary goal of visualizations is to communicate information, leading to understanding~\cite{Few2012}.
Pat Hanrahan joins the discussion and agrees with Few on the informative role of visualizations but suggests that memorable visuals can also convey information effectively. He stresses the evaluation of visualizations across various dimensions, including aesthetics, style, playfulness, and vividness. Hanrahan highlights aesthetics, noting that scientific theories suggest that attractive visuals are perceived as more usable, linking effectiveness to aesthetics~\cite{Hanrahan2007, Few2012}.

This link between effectiveness and design, but also in how far this is connected to rhetoric, is a topic that is subject to occasional discussion and theorizing. Much like rhetoric, design can be persuasive, as it often aims to influence people's thoughts, beliefs, convictions, and attitudes~\cite{Susanka2021}. Cognitive psychology today underscores the interconnectedness of reason and emotion. This reinforces the rhetoric theory that thinking and feeling are inseparable, highlighting the relationship between rationality, emotions, and trust through logos, ethos, and pathos~\cite{Kjeldsen2021}.



The potential for rhetoric as a design approach, by appealing to both rational and emotional aspects of an audience, is supported by dual-process theories. These psychological theories explain how thinking can occur in two distinct ways or as a result of two separate processes~\cite{KnapeJoachim2021PbDD}. Often, these processes involve an implicit, unconscious component and an explicit, conscious component~\cite{KahnemanDaniel2012Tfas}. Among these theories, the most relevant suggestions for visualization design are~\cite{KnapeJoachim2021PbDD}:
\begin{itemize}
    \item \textbf{Dual coding theory:} Designers should mix different codes, like languages and images, to enhance audience processing~\cite{KnapeJoachim2021PbDD}. Combining visuals and written content enables people to access information through text, visuals, or a combination of both~\cite{vavra2011visualization}. Visualization research should incorporate textual context alongside visual elements in design assessment, giving equal attention to language construction and readability to better cater to viewers' needs and enhance efficacy~\cite{stokes2021give}.
    \item \textbf{Dual system theory of cognitive processing:} To engage both rapid and reflective thinking, designers can use strategies such as connotations, associations, and metaphors~\cite{KnapeJoachim2021PbDD}. As users become more comfortable with digital tools, design principles favor simpler and standardized styles that work well across platforms like the web, tablets, and mobile devices. Despite this, many widely used visual elements, like icons and sliders, still incorporate real-world metaphors for their practical usability.
    \item \textbf{Cognitive psychological model of persuasion:} Design choices can lead to specific cognitive responses, emphasizing how information is shaped and organized in design~\cite{KnapeJoachim2021PbDD}. This model is linked to the design principle of simplicity, especially in interactive visualizations, which promotes progressive disclosure. By hiding certain elements until needed, designers can lessen cognitive overload and improve understanding.
\end{itemize}

Combining all these concepts, studies about neurological processes provide a foundation for effective communication, which is needed to deliver complex information today, especially when facing challenges like pandemics or climate change.
For science communication to effectively contribute to societal decision-making, it must bridge the gap between scientific research and the general public, presenting insights in a credible and relatable way and emphasizing their relevance to everyday life~\cite{Susanka2021}. Good data visualizations result from bringing together communication, data science, and design. When done well, they make complex datasets easier to understand and provide important insights~\cite{Islam2019}.

Despite these arguments, there is a common assumption that data collection and data visualizations are inherently objective and free from bias~\cite{Correll2019}. This perspective suggests that clear and accurate presentations of data naturally convey truth. However, this overlooks the influence of design choices and contextual framing on interpretation~\cite{boehnert2016data, Hullman2011}.
Building on this critique, we now explore the role of pathos and emotions in shaping understanding and discuss the advantages and disadvantages of emotional appeal in data visualization.

\section{Challenging Neutrality: Emotion in Data Visualizations}
\label{section:neutrality} 
To this point, the prevailing viewpoint was that data, and by extension data visualizations, were apolitical and ethically neutral, based on the belief in the inherent objectivity of data~\cite{boehnert2016data, Correll2019, akbaba2024entanglements}. Subsequently, this led to the assumption that there were no moral obligations for the designers of data visualizations in terms of collecting and visualizing data. Michael Correll states that people involved in creating data visualizations also have moral obligations and are sometimes confronted with ethical dilemmas~\cite{Correll2019}. In light of this, there has been a growing awareness in the visualization community of the need for a critical perspective that can expose underlying power relations. This perspective highlights that even seemingly objective datasets can be manipulated to favor certain points of view~\cite{Kennedy2016}. Moreover, it is essential to recognize how knowledge creation is shaped by historical and contextual factors, suggesting that data and visualizations, along with their designers and audiences, are inherently interconnected~\cite{akbaba2024entanglements}. At the same time, many visualization designers believe that individualizing and emotionalizing data visualizations can lead to a better understanding of the data, as it makes the information more accessible to the audience~\cite{Campbell2019, Kennedy2016,NormanDonaldA2003Tdoe}.
However, the principle of clarity remains an essential element for ensuring comprehensibility in data visualizations, aligning with Tufte's minimalist paradigm~\cite{Kostelnick2008}. It is challenging to deviate from these established principles and shift the emphasis from minimalism to a balanced use of logos, ethos, and pathos. From a rhetorical perspective, it is also necessary to ask what exactly is understood by clarity, for this concept is now far more complex, involving many more facets, than it did 50 years ago~\cite{Kostelnick2008}. Clarity, as per Tufte, is often associated with a high data-ink ratio and immaculate design, yet such simplifications can mask the inherent complexity of the data, presenting an image that is deceptively simpler than reality~\cite{Tufte2007, Kennedy2016}.
Therefore, using pathos techniques in presenting intricate datasets might be advantageous: ``Clarity may initially depend on pathos appeals that draw readers into the display by stirring their emotions.''~\cite{Kostelnick2008} Nonetheless, this must be carefully balanced with ethical considerations to shield the audience from malpractice~\cite{Kostelnick2008}.

While emotional appeals are often viewed as personal and contrasted with logical arguments and decision-making~\cite{KostelnickCharles2019Hvd:}, presenting information for reader comprehension is inherently subjective~\cite{Correll2019,KostelnickCharles2004MIMA}.
For instance, statistical atlases use numerous data visualizations to offer specific interpretations of the nation at particular historical moments. These visual constructs are informative and highly persuasive, functioning as implicit arguments about national development, population movements, and the integration of immigrant groups~\cite{KostelnickCharles2004MIMA}.
Rhetoric permeates all forms of communication; visual representations are shaped by distinct historical contexts, as ``ideological vacuums do not exist.''~\cite{KinrossRobin1985TRoN} In this sense, emotions enable numbers and data to be translated into forms people can relate to and understand~\cite{Bonsiepe2008, Heath2022book}. In other words: ``[I]f we don’t translate numbers into something that’s more tangible, we’re going to sacrifice in a big way.''~\cite{Heath2022}
This is why, in the following section, we discuss to what extent emotions can influence understanding and memorability.

\subsection{The Role of Emotion on Understanding \& Memorability}

One method to bridge the gap between data and human understanding is infusing data with emotional elements. Emotions catalyze action, making abstract numbers feel personal and relevant~\cite{Heath2022}. The research of Scott and Paul Slovic~\cite{SlovicPaul2007IILa, Slovic2015a} supports this position: As ``numbers get larger and larger, we become insensitive; numbers fail to trigger the emotion or feeling necessary to motivate action.''~\cite{Slovic2015vestfjaell} Aristotle and Campbell acknowledged a difference between emotional and logical strategies but considered them complementing each other rather than opposing one another~\cite{KostelnickCharles2019Hvd:}. Recent research also suggests that emotions influence understanding, transmission, and exploration of data~\cite{Wang2019}. From a cognitive science perspective, it is evident that emotions are directly linked to motivation or learning~\cite{Okabe-Miyamoto2022, TyngChaiM.2017Tioe, crossmann2007, MegaCarolina2014WMaG}. Positive emotions influence motivation and self-regulation in ways that improve academic learning, whereas negative emotions can do the exact opposite~\cite{Okabe-Miyamoto2022, MegaCarolina2014WMaG}.
Regarding emotional valence, people tend to pay more attention to emotions perceived as negative than positive~\cite{Avry2020, Zhao2022}. The negativity bias supports this assumption, stating that negative emotions attract more attention than positive ones, for example in social media, and thus generate more engagement. This is because the negative represents a potential threat---and the brain needs more cognitive resources to cope with threats~\cite{Zhao2022}.
Wang et al. argue that evoking emotions through data visualizations enhances data understanding, as emotions can aid the recipients' comprehension process~\cite{Wang2019, steigenberger2015emotions}. For these reasons, the concept of pathos has the potential to become an important component of contemporary data visualizations. As Kostelnick points out, ``[d]esigners who understand this dimension of data design can deploy technology to make their displays more engaging, humane, and usable.''~\cite{Kostelnick2016} Applying these techniques in the design process can improve emotional engagement. Building on aesthetics helps capture the audience's interest and makes data more accessible. As a result, these factors support the sense-making process and shape user expectations, leading to more effective interactions with the environment~\cite{xenakis2015aesthetics}.

Building on this, rhetoric in communication design is crucial in enabling sense-making through visual messages, fostering a shared visual language that bridges differences and reduces ambiguity. It facilitates expression and freedom and supports truth-seeking and the communication of public concerns, potentially leading to shifts in perspectives, attitudes, or behaviors by fostering social consensus~\cite{kim2010speculative}. However, it is important to recognize that the effectiveness of rhetoric comes with inherent risks and requires careful and responsible application. Misuse can occur even without intentional rhetoric, as seen in data visualization, where inaccuracies may arise from presenting incorrect data or applying misleading visual techniques~\cite{pandey2015deceptive}. Appealing to emotions poses potential downsides, which are addressed in the following section.

\subsection{Emotions and their Potential Downsides}
\label{section:dangers} 

As history has demonstrated, pathos does not only carry positive connotations. Studies have shown that the so-called ``vicarious distress'' caused by the feeling of empathy, which can be the result of pathos techniques, can influence social behavior~\cite{GrynbergDelphine2012DaVo, BatsonC.Daniel1987DaET}. In other words, if someone experiences the same suffering as another person, it might lead to defensive or blaming behavior, making it harder to engage with data visualizations. A less emotional response, however, could help make more rational decisions and assess whether the other person's suffering is something to be morally concerned about.~\cite{GrynbergDelphine2012DaVo, BatsonC.Daniel1987DaET}.

Furthermore, empathy poses a challenge due to the ``similarity bias,'' a cognitive tendency to favor individuals who share characteristics or experiences with oneself~\cite{Prinz2011a}. Neuroscientific studies reveal that empathetic reactions are stronger toward those perceived as part of one's own group~\cite{Talt2019}. This aligns with research showing that images are most powerful when they connect with preexisting values and emotions of an audience. As a result, the role of image-based rhetoric is not just to persuade but to trigger and amplify these underlying cognitive and emotional responses~\cite{Kjeldsen2021}. Cognitive psychology further confirms the deep connection between logic, emotion, and credibility, highlighting that the effectiveness of logos, pathos, and ethos in visual rhetoric depends on both the situation and the audience~\cite{Kjeldsen2021}. Educating people to think critically about images, especially those that confirm existing beliefs, remains an important process~\cite{Kjeldsen2021}. This is particularly relevant when considering how individuals may prioritize one cause or group over another based on personal connections or emotional appeals. For example, an individual may prioritize their own pet over a homeless person in a distant part of the city, or they might feel a stronger connection to a faraway country where a relative lives rather than to a neighboring nation~\cite{formsofinfluence}. 

To better understand these underlying structures, it is essential to consider the situation, facts, and cultural context. Appealing to people's values can inspire them to care deeply enough to make a difference, ultimately enriching their compassion and concern~\cite{formsofinfluence}.

In the following section, we will present George Campbell's framework for creating emotions and contextualizing them within data visualization. We will also analyze examples in Figure~\ref{fig:examples} to illustrate how these emotional appeals are used in data visualization.

\begin{table*}
\centering
  \caption{Overview of Campbell's seven circumstances and corresponding methods and empirical studies relevant to data visualizations as well as examples for each circumstance}
  \label{tab:freq}
  \begin{tabular}{llll}
    \toprule
    Campbell's circumstance & Design strategies for data visualization & Scientific literature \& studies & Visualization examples\\
    \midrule
    1. Probability & Reliable data, credibility & \cite{Kennedy2018, Rosling2019, Tufte1990, Tufte2006, Tufte2007} & \cite{MIT, noauthor_collapse_2022, Halloran}\\
   2. Plausibility & (Data) storytelling & \cite{Martinez-Maldonado2020, Sejal2019, Dahlstrom2014, NeiferThomas2020DSak, Matei2021, segel2010narrative, blount2020understanding} & \cite{pink, minard1869, Scarr, CovidDeaths2022, Halloran}\\
    3. Importance & Personalizing the data, agency & \cite{Rawlins2014, Tufte1990,  Kostelnick2016} & \cite{fung, wallach2021, martinkr, wordldatalab2021, noauthor_collapse_2022 }\\
  4. Proximity of time & Inserting prediction, temporal proximity & \cite{Campbell2019, bashir2014time} & \cite{ross2021, wordldatalab2021, meteor}\\
     5. Connection of place & Spatial proximity, role-playing  & \cite{Campbell2019, Luccioni2021, Harris2015, Lan2022} & \cite{wallach2021, wordldatalab2021, MIT}\\
    6. Relations to persons concerned & Humanizing data, anthropographics  & \cite{ dhawka2023, KostelnickCharles2019Hvd:, Morais2021} & \cite{pink, kooproutley2021, ghosh2021, Periscopic2018, Scarr}\\
  7. Interest in the consequences & Subject matter, proximity to interests & \cite{Campbell2019, Kennedy2018} & \cite{worldPopulation2015, meteor, wordldatalab2021}\\
  \bottomrule
\end{tabular}
\end{table*}

\section{Seven Circumstances by George Campbell}
\label{section:circumstances} 

We argue that the role of pathos within rhetoric and aesthetics in data visualization has not yet been clearly articulated and exemplified in the context of data visualization. Rather than questioning whether pathos should be used in general, we aim to shift the discussion to how they are already incorporated in data visualization and what effects they can have. By focusing on theories related to pathos, we aim to understand its role in enhancing audience engagement and the effectiveness of visual narratives. This analysis aims to encourage designers to critically reflect on their use of emotional elements, whether intentional or subconscious, and the broader implications it can have.
Using Campbell's seven circumstances as a framework, we will draw on theoretical insights and provide examples in Figure~\ref{fig:examples} of how practitioners implement these emotional strategies into their work.

Campbell’s ``philosophy of rhetoric'' (1776) highlights the importance of emotion and imagination in communication~\cite{Campbell2008} and states that passion (or what Aristotle called pathos) is necessary to engage with and animate ideas. Despite acknowledging the potential danger of passion, Campbell believed rhetoric should not be dismissed as an ``art of deception'' but instead be researched and used legitimately for communication~\cite{Campbell2008}. According to him, imagination makes ideas shine, memory makes them last, and emotion brings them to life~\cite{Campbell2008}.


Campbell identified seven circumstances ``that are chiefly instrumental in operating on the Passions:''~\cite{Campbell2008} \textit{Probability, plausibility, importance, proximity of time, connection of place, relations to persons concerned,} and \textit{interest in the consequences}. 
Each of these circumstances is a contextual factor that aims to heighten belief and to increase the audience's attention or sense of urgency.~\cite{Walzer1999}.
His framework~\cite{Campbell2008} has been practically adapted to data visualizations~\cite{Kostelnick2016, Campbell2019}. For this paper, we expand the framework (in Table~\ref{tab:freq}) and offer an overview of the seven circumstances, accompanied by potential design strategies for data visualization and visualization examples. Table~\ref{tab:freq} was developed through a comprehensive analysis of relevant literature, including empirical studies. While not exhaustive across all disciplines, the design strategies outlined in Table~\ref{tab:freq} reinforce the theoretical foundations presented in this paper and will be used for the analysis for Sections~\ref{section:circum1} to~\ref{section:circum7}. Each circumstance and design strategy is enriched with a) a body of scientific literature and empirical studies that examine or address these strategies and b) examples showcasing the versatility and practical application of these strategies in diverse visualizations. In this paper, we discuss six selected visualization examples (see Figure~\ref{fig:examples}) in detail, relating them to the relevant circumstances, with additional examples provided in Table~\ref{tab:freq}. In the subsequent discussion, we will examine each example's circumstances, provide reasons for these associations, and show that some visualizations employ rhetorical strategies that apply to more than one circumstance. While each circumstance targets a specific aspect of emotional engagement, their effects as well as their influence on other aspects like logos and pathos can intersect and overlap sometimes, reflecting the complex and interconnected ways in which emotions are experienced.

\subsubsection{Probability}
\label{section:circum1}
The first technique to evoke pathos is called ``probability'' and addresses the extent to which visualized data is comprehensible and tangible to the recipients~\cite{Campbell2008}.
While probability is rooted in evidence and closely tied to credibility (linked to ethos), it does have an emotional impact: Campbell explains that probability creates belief by showing data as likely or plausible, rather than certain. This sense of plausibility helps the audience feel that the data is trustworthy, which can trigger an emotional response. When data feels believable, even without being certain, it strengthens both its credibility and its emotional appeal~\cite{Campbell2008}.
Credibility is closely related to trust and is a complex and context-dependent concept~\cite{wallace2020consuming}. Importantly, credibility can coexist with uncertainty when engaging with data, and trust is further influenced by perceptions of credibility~\cite{Aristoteles}.
This aligns with Kennedy and Hill's observation that what the data themselves are about contributes to its emotional impact in visualizations~\cite{Kennedy2018}. They found that emotions towards data visualizations stemmed from various factors, including the data, design, visual style, subject matter, source, and participants' interpretation skills. Still, they noticed emotional reactions towards the data themselves, especially when the data surprised the viewers, such as alarming statistics on endangered species or the destructive power of an atomic bomb~\cite{Kennedy2018}. Understanding how credibility and emotional responses interact is important for creating visualizations that connect with audiences.
According to Rosling, the urgency instinct pushes us to react quickly to perceived threats without fully analyzing the context or probability behind the data~\cite{Rosling2019}. In the context of data visualizations, especially when alarming statistics are presented---such as those related to environmental decline or catastrophic events---viewers may feel compelled to act or form conclusions based on initial emotional responses rather than carefully assessing the credibility of the data~\cite{Rosling2019}.
For example, Figure~\ref{fig:examples}, A, developed by MIT’s Senseable City Lab, visualizes urban tree coverage across cities globally. The project shows the percentage of tree coverage (Green View Index) across different urban areas, using Google Street View (GSV) panoramas~\cite{MIT}. It is an open-source project with scientific papers available on the website, likely enhancing the credibility and rigor of the visualized data~\cite{MIT}.  For instance, while a cluttered or overly embellished graph might dramatize a steep decline in tree coverage, it risks provoking alarmist reactions rather than fostering a balanced understanding of the data. Such an approach can overshadow critical considerations, such as long-term trends or broader probabilities.  It is essential to present data clearly and efficiently, promoting understanding without distorting the message~\cite{Tufte1990, Tufte2006, Tufte2007}. This highlights the need to balance emotional resonance with well-visualized, credible data to prevent misleading interpretations and enhance the credibility of the information presented.

\subsubsection{Plausibility}
\label{storytelling}

The second circumstance is ``plausibility'': the data visualized must appear plausible and resonate with the audience. Design methods like data storytelling can help ground data within relatable contexts. By weaving data into narratives, these methods help make complex information more accessible and meaningful for lay people, who often rely on mass media to consume scientific knowledge. This favors narrative-driven formats~\cite{Dahlstrom2014}. As studies by Matei et al.\cite{Matei2021} and Neifer et al.\cite{NeiferThomas2020DSak} demonstrate, crafting resonant narratives is essential for connecting with audiences.
Research in narrative visualization design explores genres and elements that balance structured, author-driven storytelling with reader-driven interactivity~\cite{segel2010narrative}. Data story designers often use narrative patterns--repeatable templates that address common design challenges. Studies have examined which patterns are most frequently used, alone or in combination, and which ones beginners find challenging~\cite{blount2020understanding}. Storytelling approaches can also prompt deeper reflection on simulation experiences~\cite{Martinez-Maldonado2020} and make data more engaging and impactful~\cite{Sejal2019}.

One example of a practical implementation of this circumstance is seen in Figure~\ref{fig:examples}, B. This infographic uses storytelling to convey its message. It connects additional information with statistical data by guiding the viewer through a narrative that illustrates the Pink Tax's impact. This storytelling approach can make complex information more digestible and engaging, as well as highlight the systemic issues at play~\cite{pink}.
Another well-known example is Minard's Map (Figure~\ref{fig:examples}, E), showing Napoleon’s march to Moscow. The map presents two timelines: the army’s journey to Moscow and back and the corresponding temperature decline during the retreat. The alignment of the falling temperatures with the shrinking size of the army enhances the narrative, connecting it back to the human cost of military operations~\cite{minard1869}. Similarly, ``Iraq's bloody toll'' (Figure~\ref{fig:examples}, D) tells a story, though employing a somewhat different approach: It uses downward-facing, red-colored graphs that illustrate the trend in deaths over time. The downward-facing orientation of the graph creates a visual effect, resembling blood dripping and thus conveying the impression of continuous loss and mortality~\cite{Scarr}.


\subsubsection{Importance}

\begin{figure*}
	\centering
	\setlength{\fboxsep}{1pt}
	\setlength{\fboxrule}{1pt}
	
 \includegraphics[width= 515pt]{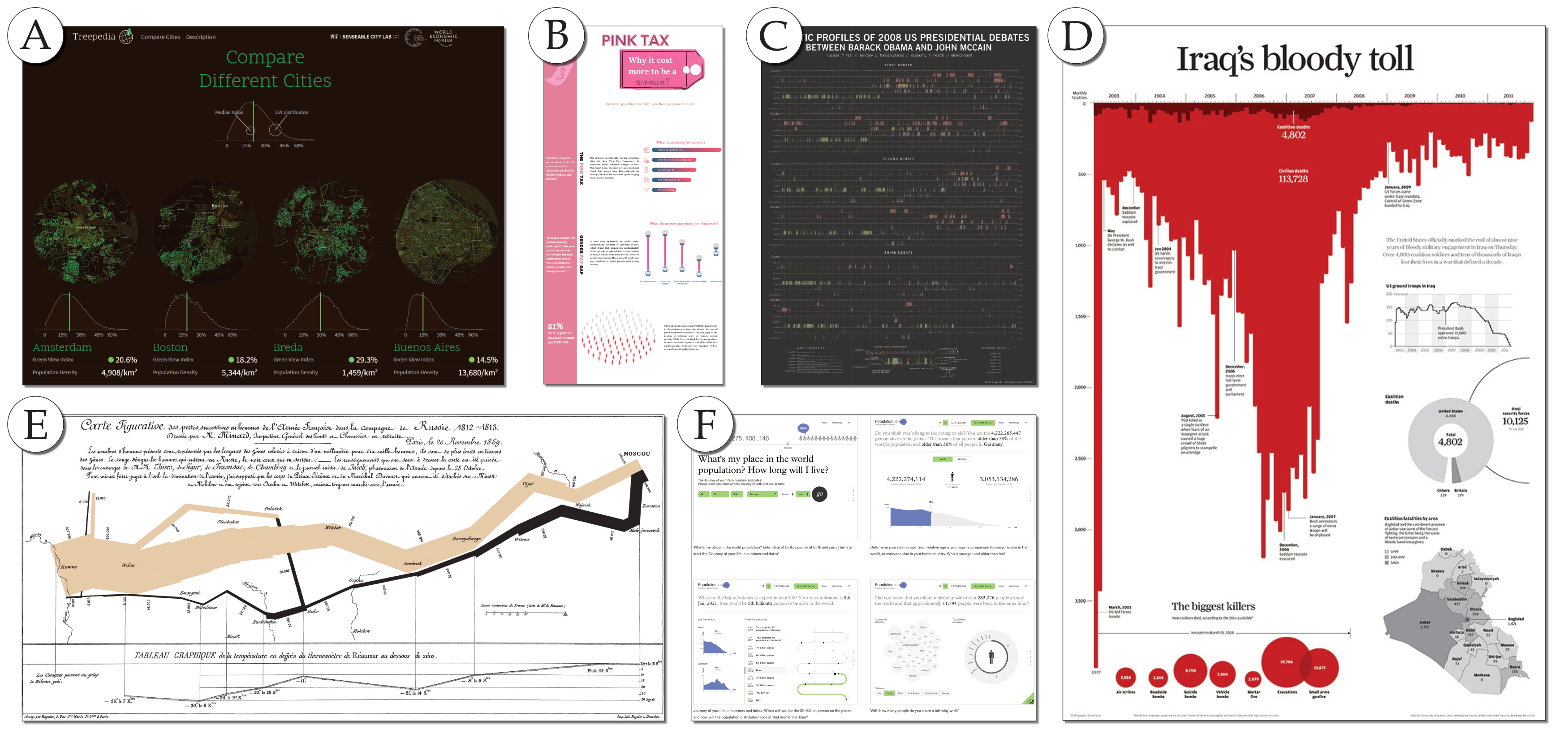}
	\caption{These six visualizations demonstrate how pathos techniques are implemented in data visualizations. They cover a range of topics, from global issues like urban greenery (A)~\cite{MIT}, and gender-based pricing disparities (B)~\cite{pink}, to a thematic analysis of the 2008 U.S. presidential debates (C)~\cite{martinkr}, and the human toll of the Iraq war (D)~\cite{Scarr}. Additionally, they include visualizations of historical events like Napoleon's march to Moscow (E)~\cite{minard1869}, as well as a contemporary personalized population calculator that connects global statistics to individual lives (F)~\cite{wordldatalab2021}. For full-scale images, please refer to the supplemental material. \textit{Visualizations A, B, C, D and F are reproduced with permission, visualization E is in the public domain \ccPublicDomain}.}
	
	\label{fig:examples}
\end{figure*}

In this context, the term ``importance'' suggests that people are likely to have a stronger emotional reaction when certain elements, such as people, places, or events, are meaningful to them~\cite{Campbell2008, Campbell2019}. Giving individuals agency in how they engage with data can enhance the sense of importance, allowing them to connect more deeply to the information presented. The growing global accessibility of the web and data has increased the variety of data visualizations available to viewers~\cite{Rawlins2014}. Digital visualizations, for example, can enable interaction, aligning with Tufte’s vision~\cite{Tufte1990} that they should allow viewers to select, narrate, reinterpret, and personalize data~\cite{Rawlins2014}. Interactivity allows viewers to personalize the data, emphasizing what is important to them while hiding what is not. It can enable personalizing the display itself, including chart types and their visual elements (like background color)~\cite{Kostelnick2016}. 
Figure~\ref{fig:examples}, F, ``population.io'', shows an interactive visualization, calculating a viewer's position within the global population and lifespan based on their respective demographic factors. The creators believe that ``demographic data can play an important role in understanding the social and economic developments of our time.''~\cite{wordldatalab2021}
Another example of a contemporary approach to this circumstance is Figure~\ref{fig:examples}, C. It shows a thematic analysis of the presidential debates in 2008 between Obama and McCain. The debates are broken down into three separate timelines, each representing one of the debates, allowing viewers to track how the conversation shifted over time. The direct comparison between Obama's and McCain's thematic foci during the debates helps the viewer to see how much time was spent on themes that are important to themselves and illustrates the differing priorities of each candidate~\cite{martinkr}.

\subsubsection{Proximity of Time}

This technique is based on the idea that people and events in temporal proximity to the audience create a stronger emotional connection and reaction~\cite{Campbell2019, Campbell2008}. The corresponding design strategy for data visualizations involves inserting predictions and emphasizing temporal proximity to enhance affective impact (see Table~\ref{tab:freq}). In a 2019 study, Campbell and Offenhuber applied this technique using a shelter animal dataset, incorporating predictions of imminent death to create a sense of urgency. The findings revealed heightened emotional responses among participants, particularly in terms of interest, disgust, fear, and disappointment, with ``interest'' being the emotion most significantly influenced by the temporal proximity technique~\cite{Campbell2019}. Similarly, another study on social cognition examined how subjective experiences of temporal proximity can increase motivation for long-term collective goals, finding that when participants perceived distant future climate change consequences as temporally closer, their pro-environmental motivation and behavior improved~\cite{bashir2014time}.
An example for proximity of time is ``population.io'' (Figure~\ref{fig:examples}, F). By using real-time data, it conveys a sense of urgency and importance, and by visualizing the viewer's lifespan (statistically) and how they compare to others worldwide~\cite{wordldatalab2021}.

\subsubsection{Connection of Place}
This circumstance is based on the theory that people or events that have local proximity evoke stronger emotions~\cite{Campbell2019}. A design strategy for data visualization is creating a spatial proximity (as outlined in Table~\ref{tab:freq}), which can involve interactive experiences, such as the selection of a location by the viewer. For instance, Luccioni et al.~\cite{Luccioni2021} used an AI-based approach to visualize the impacts of climate change. They enabled viewers to choose a location and generate a visualization of the place's transformation over 50 years due to climate events. This strategy creates proximity by allowing personal selection of the place, an thus enhancing personalization and concreteness~\cite{Luccioni2021}. Lan et al. highlight the significance of connecting viewers to the real-life scenes and characters represented by data through \textit{role-playing}. In their study, they encouraged participants to immerse themselves in a specific scenario---such as playing the role of a lifesaver---to experience firsthand the challenges faced by others, like the complexities of delivering first aid in particular communities~\cite{Lan2022}. By placing participants within these scenarios, this approach fosters a ``connection of place,'' grounding the data in the tangible realities of the settings and communities involved. They found, that this approach was able to enhance emotional engagement and improve long-term recall~\cite{Lan2022}. 
Similarly, Harris emphasizes the importance of incorporating both near and far perspectives in data visualization, suggesting that focusing on local data alongside broader trends can evoke stronger emotional responses~\cite{Harris2015}.
Treepedia (Figure~\ref{fig:examples}, A) fits this context, as it displays tree coverage across cities worldwide, from Amsterdam to Xapalta, making it geographically relevant to a broad audience. The ability to zoom in on specific cities provides a localized perspective, creating a stronger connection between the place and the viewer~\cite{MIT}.

\subsubsection{Relations to Persons Concerned}

According to Campbell's theory, emotional responses are more likely to arise when the information involves individuals connected to the target audience. He emphasizes the power of personal connections, stating, ``it is the persons, not the place, that are the immediate objects of the passions love or hatred, pity or anger, envy or contempt.''~\cite{Campbell2008}


In data visualization, humanizing the data is one design strategy. Isotypes---standardized visual symbols or pictograms used to convey information---and anthropographics, which are human-centered visual representations, can make data more relatable. Including human figures, whether overt or subtle, elicits emotional responses~\cite{KostelnickCharles2019Hvd:}. There is growing research in anthropographics (emphasizing the human importance of datasets), which explores how visualization designers and data journalists use design strategies to help audiences connect with and better understand data about people~\cite{dhawka2023, Morais2022}. Although visual representations of bodies can appeal to people's emotions, those are often subjective and difficult to measure, despite their physical effects~\cite{KostelnickCharles2019Hvd:}.
Figure~\ref{fig:examples}, B shows an example of using human shapes. It incorporates human-shaped figures to represent women, illustrating the disparities in spending related to the ``Pink Tax''. These figures are used, in this case, to make the economic implications of the Pink Tax more relatable~\cite{pink}. Figure~\ref{fig:examples}, D, ``Iraque's bloody toll,'' also establishes a personal connection with those affected. By focusing on individual data points, such as the leading causes of death and U.S. military casualties, the visualization evokes an emotional response, emphasizing the human impact of the war~\cite{Scarr}.

\subsubsection{Interest in the Consequences}
\label{section:circum7}
``Of all the connective circumstances, the most powerful is interest,'' asserts George Campbell. The intensity of people's interest creates a sense of proximity, fostering empathetic feelings~\cite{Campbell2008}.

Research indicates that the ``subject matter''~\cite{Kennedy2018} significantly influences emotional responses when engaging with data visualizations, encompassing emotions such as anger, sadness, relief, and empathy. However, if viewers have strong negative emotions toward the subject matter, such as hatred or anger, this can hinder their engagement with the data~\cite{Kennedy2018}. Campbell \& Offenhuber's study, using a dataset on animal shelters, also exemplifies this concept. Before seeing the visualization, participants could decide which data they wanted to see, they could choose between ``dogs'' or ``cats.'' The results validated ``the logical assumption that people are more interested in the visualization when the data aligns with their interests,''~\cite{Campbell2019} reinforcing the importance of aligning visualizations with audience interests.
``Population.io'' (Figure~\ref{fig:examples}, F) exemplifies this circumstance, as it inherently plays on viewers' interest in the consequences of demographic trends for their own lives: how long they might live, what major population milestones they will witness, and how their position in the world population changes over time, potentially influencing personal decisions about health, lifestyle, and future planning.~\cite{wordldatalab2021}.

\section{Literature-Informed Design Suggestions for Rhetorical Strategies}
\label{section:design_suggestions} 

\begin{table*}
\centering
\renewcommand{\arraystretch}{1.5}
\caption{Design suggestions for effective data visualizations drawn from the reviewed literature}
\begin{tabular}{|p{5cm}|p{12cm}|}

\hline

\textbf{Design Suggestions} & \textbf{Description} \\ \hline
Prioritizing clarity \textit{(logos)} & Visualizations should prioritize clarity and readability to ensure the data is easily understandable. Avoid unnecessary embellishments that may distract from the core message. \\ \hline
Enhancing credibility \textit{(ethos)} & Clearly communicate data sources and methodologies used. Provide information about the expertise and qualifications of the visualization designer or team. \\ \hline
Appealing to emotions \textit{(pathos)} & Focus on aesthetics, including color schemes, typography, and layout. Experiment with visual styles to align with the intended emotional tone. \\ \hline
Personalizing the data \textit{(relations to persons concerned)} & Humanizing data by incorporating personal stories, testimonials, or case studies. Use anthropographics, icons or isotypes to resonate with the audience's experiences and perspectives. \\ \hline
Balancing information and simplicity \textit{(probability, plausibility)} & Presenting data in a balanced manner, avoiding complexity or oversimplification. Utilize data storytelling techniques for a plausible and relatable presentation. \\ \hline
Considering cultural and ethical contexts \textit{(importance, interest in the consequences)} & Designing visualizations that are sensitive to perspectives and values of the audience, while considering what matters most to them. By highlighting the implications or consequences of the data, visualizations can better resonate with viewers and connect to their interests. \\ \hline
Emphasizing temporal and spatial aspects \textit{(proximity of time, connection of place)} & Leveraging temporal and spatial elements such as timelines, maps, and interactive features to provide context and improve understanding. \\ \hline
Incorporating optional interactive features & Fostering viewer engagement by allowing data exploration. Empowering the audience to derive insights based on personal interests. \\ \hline
\end{tabular}

\label{tab:design_suggestions}
\end{table*}

This paper presents an extensive body of literature on rhetorical strategies for data visualization design. Drawing on these insights, Table~\ref{tab:design_suggestions} consolidates practical strategies that aim to enhance visualizations' clarity, emotional engagement, and overall effectiveness based on the literature presented. These strategies balance rhetorical elements---logos (clarity), ethos (credibility), and pathos (emotional appeal), with aesthetic and functional considerations. 

However, designers must navigate the inherent trade-offs between these elements. For instance, while incorporating human figures can increase emotional resonance and improve thematic understanding of complex topics like immigration or unemployment, it may also introduce perceptual challenges that reduce clarity. This conflict between clarity and emotional engagement is common in visualization design. As Kostelnick notes, designers continue to use figural forms ``because they recognized their emotional value through audience engagement, identification with the figures and [...] emotional responses''~\cite{KostelnickCharles2019Hvd:}. 

To resolve these conflicts, designers can apply a balanced approach, aiming for emotional elements to complement rather than compromise clarity. For example, annotations, fitting color schemes, and thoughtful use of figurative representations and embellishments can help mitigate potential confusion while engaging the audience emotionally.
By considering the cultural and ethical background of the audience and highlighting the relevance of the data that may interest them, designers can create visualizations that resonate on a personal level and connect the data and the audience.

The design suggestions presented provide an alternative to traditional approaches to creating data visualizations. Still, it is important that designers find a balance between these design suggestions to foster emotional engagement on the one hand but still maintain basic principles like clarity on the other hand. We do see that this balance is a subjective act. Still, when done well, it can help designers create visualizations that effectively convey complex data while being accessible and engaging to a certain audience.

\section{Discussion \& Key Takeaways}
\label{section:limitations} 

History reveals that emotions have consistently played a role in communication and rhetoric. Recent interest, highlighted by Lee-Robbins \& Adar, recognizes the effectiveness of affective learning and pathos techniques in data visualizations to capture attention and induce behavior change~\cite{lee-robbins_affective_2022}. We argue that designers of data visualizations bear the responsibility of intentional design, using rhetorical strategies to communicate data effectively. Our paper clarifies fundamental terminology and offers working definitions within the context of data visualization and stresses the deliberate application of rhetoric, emphasizing the power of visualizations to influence cognitive and affective dimensions. Tailoring persuasive strategies to the intended audience is crucial to effective (scientific) communication, making complex information more accessible and understandable. While a careful approach is necessary in the scientific domain, effective communication methods can enhance comprehension and reduce cognitive load.

Essentially, our paper provides a synthesis of perspectives throughout history, including the following takeaways for the use of data visualization for diverse audiences:
\begin{itemize}
\item \textbf{Offering working definitions of rhetoric, pathos, and aesthetics in data visualizations:} Our research draws from a broad set of literature from diverse scientific fields and defines the different terms aesthetics, rhetoric, and pathos and puts these concepts into the context of data visualizations. Offering working definitions for visualization researchers is a first step in fostering a shared understanding within the field, reducing ambiguity, and enabling effective communication among researchers, designers, and practitioners.

\item \textbf{Etymological understanding of rhetoric, pathos, and aesthetics:} Tracing rhetorical strategies, especially pathos, back to Aristotle, we highlight the historical context to guide effective use in contemporary data communication. By exploring George Campbell's influential rhetorical framework, we analyze the lasting relevance of historical theories.

\item \textbf{Ubiquity of emotions in communication:} Recognizing emotions as part of communication, we note that people naturally infuse emotions into data visualization. Often, this occurs unknowingly, highlighting the need for a holistic understanding of the possible implications of using pathos/emotions in data visualization design.

\item \textbf{Emotional appeal can be beneficial for data visualizations:} Our research underscores the significance of emotional appeal (pathos) in data visualization, asserting that emotions are essential for connecting, understanding, and engaging with complex information. However, there are real-world trade-offs using pathos/emotional appeals in data visualization design.

\item \textbf{Challenging neutrality of data:}  We challenge the perceived neutrality of data by contextualizing ethical concerns within emotional and pathos-driven appeals. Our research highlights the link between aesthetics and the effectiveness of data visualization, suggesting that well-designed visualizations enhance usability and play a role in sense-making. Addressing the neglect of emotional appeals, our research poses a central question on how understanding the dynamics of emotion, rhetoric, and aesthetics can improve ethical and effective visual rhetoric in diverse contexts.

\end{itemize}

\subsection{Limitations and Future Work}

There are other techniques to appeal to emotions, which are only partially mentioned in this paper, such as: visual imagery embellishments or art (e.g., \cite{bateman_useful_2010, lan_smile_2021}), narratives or stories (e.g., \cite{Kostelnick2016, packer_memorybook_2012, Hullman2011, lee_sketchstory_2013}), sensitive or sensory experiences (e.g., \cite{Elli}), immersive data visualizations (e.g., \cite{ivanov_walk_2019, romat_dear_2020, donalek_immersive_2014}) and animation (e.g., \cite{lan_kineticharts_2022}). Also, framing effects in narrative visualizations can potentially create more engaging visualizations~\cite{Hullman2011}. While it is possible to find connections to Campbell's seven circumstances in these areas, this was outside the scope of this work because it would have extended our intended limits, potentially undermining the clarity and depth of our main findings.
Although extensive research has been conducted, relevant papers may have been missed due to varying keywords across different scientific disciplines. To demonstrate that rhetorical strategies are used in past and contemporary visual communication, even if they are not explicitly labeled, this paper presents theories, studies, and examples to highlight the ubiquity of rhetoric and, by extension, emotion and aesthetics in this field.
However, the visualizations mentioned as examples are only illustrative.
Future research building on this paper could include:
\begin{itemize}
  \item Investigating emotional appeals in data visualization to understand their influence on understanding the data, memorability, decision-making, and behavior change.
  \item Exploring the role of aesthetics in data visualizations and how it affects viewer engagement.
  \item Examining specific visual rhetoric strategies for promoting desired behaviors through data visualization.
  \item Conducting longitudinal studies to assess the sustained impact of data visualization with particular rhetorical elements on behavior change and memorability.
  \item Deepening the exploration of aesthetic concepts unique to data visualization, distinct from traditional concepts such as pathos.
\end{itemize}

\section{Conclusion}
\label{section:conclusion} 
To holistically interpret and evaluate data visualizations, it is crucial to first understand the rhetorical strategies embedded in their design. This understanding helps designers and researchers recognize how visualizations shape perception and convey intended messages. Influenced by Tufte, the visualization community has traditionally emphasized minimalism and clarity, prioritizing clear data communication over aesthetic appeal. However, there has been a recent shift towards a greater consideration of rhetorical elements and the communicative aspects of data visualizations.


This work aims to create awareness of etymological (examining the origins and evolution of concepts and terminology) and epistemological (exploring how knowledge is formed and validated) factors within the visualization community. By integrating these perspectives, we seek to expand the field beyond conventional approaches and challenge established beliefs. To the best of our knowledge, this is the first work to provide working definitions of rhetoric, pathos, and aesthetics as holistic concepts in data visualization.
Recognizing these foundational rhetorical concepts and understanding their etymological evolution opens new perspectives on how visual rhetoric is addressed in academic discourse and offers directions for future research in this area. Whether we are aware of it or not, rhetoric and aesthetics are omnipresent in scientific communication. This paper aims to create a deeper understanding of these concepts in relation to data visualizations and their impact on an audience. This can help designers create more targeted visualizations, avoid unintended effects, and communicate more effectively. Expanding the scientific discourse to include techniques that evoke pathos enhances data visualization research and the creation of more effective visualizations.


\bibliographystyle{IEEEtran}
\bibliography{ref.bib}

\end{document}